# On neutrino family problem

## Anatoli Vankov

(Department of Physics, Eastern Illinois University, cfaav@eiu.edu)

The experiment conducted at the Brookhaven AGS at 1962 [1] on the interaction of high-energy neutrinos with matter is considered in literature the observation of electron and muon types of neutrinos. After analysis of this experiment we have concluded that the latter does not give direct evidence of the neutrino family existence.

It is thought that three charged leptons (electron, muon, tau) and three neutral ones (electron, muon and tau neutrinos) exist in nature. In Particle Physics they are classified as three doublet families, or generations: ($e^-, \nu_e$); ($\mu^-, \nu_\mu$); ($\tau^-, \nu_\tau$). Correspondingly, there are three similar families of anti-leptons. This lepton concept does not have a deep physical ground but it matches well the three quark families: (*up, down*); (*charm, strange*); (*top, bottom*) with their anti-quark families. Such particle classification in the Standard Model has existed for a long time not only for esthetic reason but also due to experimental evidence of an apparent physical distinction between electron and muon neutrino types. While the tau-neutrino is mostly known theoretically, neutrinos associated with the electron and muon have been widely studied experimentally. For example, the neutrino mass was experimentally estimated from beta-decay process, in which a neutron transforms into a proton with emission of an electron and antineutrino:

$$n \rightarrow p + e^- + \bar{\nu} \qquad (1)$$

Historically a hint of a physical difference between neutrinos in different reactions came from the fact that in muon decay the electromagnetic transition

$$\mu \rightarrow e + \gamma \qquad (2)$$

is not observed. Instead a neutrino and an antineutrino are emitted:

$$\mu \rightarrow e + \nu + \bar{\nu} \tag{3}$$

For this and some other reasons it was suggested that there are at least two types of neutrinos, $\nu_e$ and $\nu_\mu$, so, in reactions (1) and (2) we should put $\nu_e$, and in reaction (3) neutrinos should be specified as $\nu_e$ and $\nu_\mu$, correspondingly. To validate this assumption a "two-neutrino" experiment was conducted in 1961-1962 [1]. A large flux of neutrinos and antineutrinos was generated at the Brookhaven AGS Facility. Neutrinos were secondary particles in the decay of pions into muons and were therefore presumably of the muon type. A thick iron shield filtered out the muons to leave "pure" neutrinos in the beam. In this experiment neutrinos and antineutrinos were not separated. A primary proton beam was pulsed, and each pulse triggered a registration system, which was characterized by some limited time resolution minimizing counts of "unwanted" events.

Experimenters were looking for reactions:

$$\nu_\mu + n \rightarrow \mu^- + p \tag{4}$$

$$\bar{\nu}_\mu + p \rightarrow \mu^+ + n \tag{4a}$$

and

$$\nu_e + n \rightarrow e^- + p \tag{5}$$

$$\bar{\nu}_e + p \rightarrow e^+ + n \tag{5a}$$

It was expected that if there were no distinction between electron and muon types of neutrino then reaction (4), (4a) and (5), (5a) would occur with equal likelihood. Otherwise only reaction (4), (4a) involving the muon type of neutrino would be observed, and this was exactly the case. After the experiment was finished and experimental data were analyzed the experimenters found that a majority of observed events were of "muon-type". They concluded that there was a physical distinction between neutrinos of "muon type" and "electron type". The experiment utilized state-of-the-art technology of the time, and its results were classified as a new discovery in Physics. It was appropriate that the authors were awarded a Nobel Prize. Other neutrino experiments conducted later were either of the same type or inconclusive about the neutrino family concept. Hence, a question arises as to how strong the experimental evidence is in support of this concept.

In the author's opinion, this experiment does not seem conclusive in confirming the concept. Experimenters showed that muons were indeed produced in reaction (4), (4a) in a flux of neutrinos of "muon type". A similar experiment with neutrinos of "electron type" has not been performed and hardly could be conducted at all. Had the experiment been done, it could have resulted in (5), (5a) (*a confirmation of the neutrino family concept*) or (4), (4a) (*no proof of the concept*).

The second option has not be excluded and should be investigated in a context of the absence of the electromagnetic mode of muon decay (2). We assume that all neutrinos are massless, identical, and left-handed, as was suggested in the Glashow-Weinberg-Salam Electroweak Theory. Some new property of matter has to be considered, which is illustrated in the following example.

Imagine a reaction of a neutron production under conditions of super-high gravitational pressure, which takes place, probably, in a neutron star:

$$p + e^- \rightarrow n + \nu \qquad (6)$$

The penetration of an electron into the region enveloped by the first Bohr orbit requires energy for overcoming a rising force due to degeneracy pressure. A proper mass of an electron has to increase in a process of climbing the barrier until it reached a value of the muon proper mass (the phenomenon of a proper mass variation is discussed in detail in [2]). At the *muonium* compound stage the muon should be considered the weakly bound excited electron, which is subject to a spontaneous decay. Our hypothesis is that in so called weak interactions quantum-mechanical systems are formed, in which transitions between states of excitation are realized through a process of resonance absorption and emission of neutrinos and antineutrinos. Hence, a selection rule should include $\Delta S = \pm \frac{1}{2}$. In the above example the compound system "proton-muon" results in a final "neutron" state with given electric and magnetic moments, and dynamics of the system is not known in detail. There could be reasons for electromagnetic transitions to be suppressed when a neutrino channel is open.

Thus, the neutron was formed as a result of absorption by the proton of the electron coupled with an antineutrino after a neutrino has flown away. It is clear that the neutron is a particle having "right-handedness" in excess by nature, as was found in many observations of P-violation. On the other hand, the system "neutron plus neutrino at large" has a perfect parity symmetry. Hence, there is no reason to worry about the left-right asymmetry of the world we live in. The reaction (6) should be expressed in a more detailed form:

$$p + e^- \rightarrow p + \mu^- \rightarrow p + \{e^- + \bar{\nu}\} + \nu \rightarrow n + \nu \qquad (7)$$

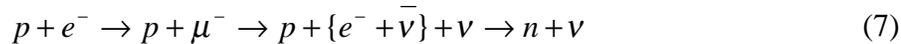

Our thought is that in the experiment [1] a stimulated neutron decay (4) (a reverse of (7)) has been observed:

$$\nu + n \to \mu^- + p \qquad (7a)$$

The similar scheme is appropriate for the reaction (4a). In (7) and (7a) neutrinos are not family labeled: the reaction (7a) should take place regardless of a neutrino source. Reaction (5) does not exist. Instead, the products of muon decay should be observed in a time-delay pulse regime (again regardless of a neutrino source):

$$\nu + n \to p + e^- + \{\nu + \overline{\nu}\} \qquad (7b)$$

It seems that a small percentage of contribution (7b) (the ratio of pulse time interval of detector openness to the muon lifetime) has been actually observed.

The suggested neutrino property of facilitation in matter interaction means the existence of a universe neutrino image in emission and absorption lines. If so, a neutrino spectroscopy is possible in principle.